\documentstyle[pstricks,pst-plot]{espart}

\begin{document}

\begin{frontmatter}

\title{Effective interactions and shell model studies of heavy tin isotopes}

\author[oslo]{A.\ Holt},
\author[oslo]{T.\ Engeland}, 
\author[nordita]{M.\ Hjorth-Jensen}, and
\author[oslo]{E.\ Osnes}
\address[oslo]{Department of Physics,
         University of Oslo, N-0316 Oslo, Norway}
\address[nordita]{Nordita, Blegdamsvej 17, DK-2100 K\o benhavn \O, Denmark}

\maketitle

\begin{abstract}
We calculate the low-lying spectra of heavy tin isotopes from $A=120$ to 
$A=130$ using the $2s1d0g_{7/2}0h_{11/2}$ shell to define the model
space. An effective interaction has been derived using $^{132}$Sn as
closed core  employing perturbative many-body techniques.
We start from a nucleon--nucleon potential derived from modern meson exchange
models. This potential is in turn renormalized for the given medium, 
$^{132}$Sn, yielding the nuclear reaction matrix, which is then used in 
perturbation theory to obtain the shell model effective interaction.
\end{abstract}

\end{frontmatter}

\section{Introduction}

The tin isotopes offer a unique opportunity for examining the
microscopic foundation of various phenomenological nuclear
models. In the Sn isotopes ranging from mass number $A=100$ to
$A=132$, neutrons are filling the subshells between the magic
numbers 50 and 82, and thus it is possible to examine how well
proton-shell closure at mass number 50 is holding up as valence
neutrons are being added, how collective features are developing,
the importance of certain many-body effects, etc.
Though, one of the problems in theoretical calculations
of properties of light tin isotopes is the fact that the 
single-particle energies of $^{101}$Sn are not known. 
This reduces the predictive power of theoretically calculated
interactions to be used in the spectroscopy of light tin isotopes.
For $^{131}$Sn however, the single-particle energies are known \cite{bf84},
a fact which allows one to discriminate between various effective 
interactions. 

Recently, several
theoretical results have been presented for the light tin isotopes
\cite{ehho95,ehho93,heho94a,hko95,nicu94,brown94}, however, for the 
heavy tin isotopes, only few 
microscopic calculations are available.
In Ref.\ 
\cite{nicu93}, Insolia {\em et al.} study the spectra of odd and even 
isotopes in the framework of a multistep 
shell model BCS formalism,
using matrix elements for the effective interaction extrapolated
from those in the lead mass region \cite{pomar90}. 

The aim of this work is to derive a more appropriate 
effective interaction for heavy tin isotopes and perform
extensive shell model studies of the isotopes from
$^{120}$Sn to $^{130}$Sn.
The effective neutron-hole interaction
is calculated with respect to
$^{132}$Sn as a closed shell core, with a model space which 
includes the orbitals
$2s_{1/2}$, $1d_{5/2}$, $1d_{3/2}$, $0g_{7/2}$ and $0h_{11/2}$.
The perturbative many-body scheme employed to calculate such
an effective interaction starts with the free nucleon-nucleon
interaction. This interaction is in turn renormalized
taking into account the specific nuclear medium. The medium
renormalized potential, the so-called $G$-matrix, is then
employed in a perturbative many-body scheme, as detailed in
Ref.\ \cite{hko95} and reviewed in the next section. 

This work falls in four sections. In section 2 we first describe how to 
calculate an effective interaction appropriate for heavy tin isotopes, 
using perturbative many-body techniques, and at the end give a brief 
presentation of the basis for the shell model calculation.
The discussion of the results for both even and odd isotopes 
are given in section 3, and concluding remarks are drawn in section 4. 

\section{Effective interaction and shell model calculations}

The aim of microscopic nuclear structure calculations is to derive
various properties of finite nuclei from the underlying 
hamiltonian describing the interaction between 
nucleons. 
When dealing with nuclei, such as the heavy tin isotopes with $A\sim 132$, 
the full dimensionality of the 
many-body Schr\"{o}dinger equation for an $A$-nucleon system
\begin{equation}
     H\Psi_i(1,...,A)=E_i\Psi_1(1,...,A),
     \label{eq:full_a}
\end{equation}
becomes intractable and one has to seek 
viable approximations to Eq.\ (\ref{eq:full_a}). 
In Eq.\ (\ref{eq:full_a}), $E_i$ and $\Psi_i$ 
are the eigenvalues and eigenfunctions
for a state $i$ in the Hilbert space.

\begin{figure}[htbp]
\setlength{\unitlength}{1.4cm}
\begin{center}
\begin{picture}(2,4)(0,-1)
\newcommand{\lc}[1]{\put(0,#1){\line(1,0){1}}}
\newcommand{\ls}[2]{\put(2,#1){\makebox(0,0){{\scriptsize $#2$}}}}
\newcommand{\lsr}[2]{\put(2,#1){\makebox(0,0){{\scriptsize $#2$}}}}
\put(-.25,3.4){\makebox(0,0){\large MeV}}
\thicklines
\put(-.75,-.5){\line(0,1){4}}
\multiput(-.75,.0)(0,1){4}{\line(1,0){.1}}
\multiput(-.75,.5)(0,1){3}{\line(1,0){.05}}
\put(-0.55,3){\makebox(0,0){3}}
\put(-0.55,2){\makebox(0,0){2}}
\put(-0.55,1){\makebox(0,0){1}}
\put(-0.55,0){\makebox(0,0){0}}
\lc{0.000}   \ls{0.000000}{3/2+ \;\;0.000}
\lc{0.2418}   \ls{0.2418}{11/2- \;\;0.242}
\lc{0.3316}   \ls{0.416}{1/2+ \;\; 0.332}
\lc{1.6546}   \ls{1.6546}{5/2+ \;\;1.655}
\lc{2.4340}   \ls{2.4340}{7/2+ \;\;2.434}
\put(0.5,-.3){\makebox(0,0){{\large $^{131}$Sn}}}
\end{picture}
\end{center}
\caption{Experimental single-hole energies for the orbits
$2s_{1/2}$, $1d_{5/2}$, $1d_{3/2}$, $0g_{7/2}$ and $0h_{11/2}$
in $^{131}$Sn.}
\label{fig:fig1}
\end{figure}
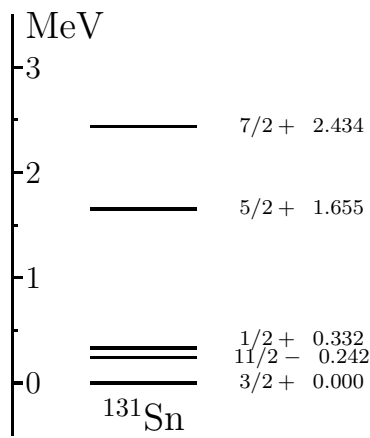

In nuclear structure calculations, one is normally 
only interested in solving Eq.\ (\ref{eq:full_a})
for certain
low-lying states. It is then customary to divide the Hilbert space
in a model space defined by the operator $P$
\begin{equation}
     P=\sum_{i=1}^{d}\left | \psi_i\right\rangle 
     \left\langle\psi_i\right | ,
\end{equation}
with $d$ being the size of the model space, and an excluded space
defined by the operator $Q$
\begin{equation}
     Q=\sum_{i=d+1}^{\infty}\left | \psi_i\right\rangle 
     \left\langle\psi_i\right | ,
\end{equation}
such that $PQ=0$.
The assumption then is that the components of these low-lying
states can be fairly well reproduced by configurations consisting
of a few particles and holes occupying physically selected orbits.
These selected orbitals define the model space. In the present 
work, the model space to be used both in the shell model calculation
and the derivation of the effective interaction is given by the
orbitals
$2s_{1/2}$, $1d_{5/2}$, $1d_{3/2}$, $0g_{7/2}$ and $0h_{11/2}$.
The single-particle energies, taken from Ref.\ \cite{bf84}, are displayed 
in Fig.\ \ref{fig:fig1}.
\\
\\

Eq.\ (\ref{eq:full_a})  can then be rewritten as a secular equation
\begin{equation}
    PH_{\mathrm{eff}}P\Psi_i=P(H_{0}+V_{\mathrm{eff}})
    P\Psi_i=E_iP\Psi_i,
\end{equation}
where $H_{\mathrm{eff}}$  now is an effective hamiltonian acting solely
within the chosen model space. The term $H_0$
is the unperturbed hamiltonian while the effective interaction is
given by
\begin{equation}
  V_{\mathrm{eff}}=\sum_{i=1}^{\infty} V_{\mathrm{eff}}^{(i)},
\end{equation}
with $ V_{\mathrm{eff}}^{(1)}$,  $ V_{\mathrm{eff}}^{(2)}$,
 $ V_{\mathrm{eff}}^{(3)}$ etc.\ being effective one-body, two-body,
three-body interactions etc. 
It is also customary in nuclear shell model calculations to add
the one-body effective interaction  $ V_{\mathrm{eff}}^{(1)}$
to the unperturbed part of the hamiltonian so that
\begin{equation}
    H_{\mathrm{eff}}=\tilde{H}_{0}+  V_{\mathrm{eff}}^{(2)}+
    V_{\mathrm{eff}}^{(3)}+\dots,
\end{equation}     
where $\tilde{H}_{0}=H_{0}+V_{\mathrm{eff}}^{(1)}$. This allows us,
as in the shell model, to replace the eigenvalues of 
$\tilde{H}_{0}$ by the empirical single--particle energies 
for the nucleon orbitals of our model space, or valence space, e.g.,
$2s_{1/2}$, $1d_{5/2}$, $1d_{3/2}$, $0g_{7/2}$ and $0h_{11/2}$, the
valence neutron holes with respect to $^{132}$Sn. 
Thus, the remaining quantity to calculate is the two- or more-body
effective interaction 
$\sum_{i=2}^{\infty} V_{\mathrm{eff}}^{(i)}$.
In this work we will restrict our attention to the derivation of
an effective two-body interaction 
\begin{equation}
      V_{\mathrm{eff}}=V_{\mathrm{eff}}^{(2)},
\end{equation}
using the many-body methods discussed in Ref.\ \cite{hko95}
and reviewed below.
The study of effective three-body forces will be deferred to a later
work \cite{eh97}. 

Our scheme to obtain an effective hole--hole interaction appropriate for 
heavy tin isotopes
starts with a free nucleon--nucleon  interaction $V^{(2)}$ which is
appropriate for nuclear physics at low and intermediate energies. 
At present, there are several potentials available. The most recent 
versions of Machleidt and co--workers
\cite{cdbonn}, the Nimjegen group \cite{nim} and the Argonne
group \cite{v18} have a $\chi^2$ per datum close to $1$.
In this work we will thus choose to work with the charge--dependent
version of the Bonn potential models, see Ref.\ \cite{cdbonn}.
The potential model of Ref.\ \cite{cdbonn} is an extension of the 
one--boson--exchange models of the Bonn group \cite{mac89}, where mesons 
like $\pi$, $\rho$, $\eta$, $\delta$, $\omega$ and the fictitious
$\sigma$ meson are included. In the charge-dependent version
of Ref.\ \cite{cdbonn}, the first five mesons have the same set
of parameters for all partial waves, whereas the parameters of
the $\sigma$ meson are allowed to vary. 

The next step 
in our perturbative many--body scheme is to handle 
the fact that the repulsive core of the nucleon--nucleon potential $V$
(herafter, we let $V$ stand for the nucleon--nucleon potential
$V^{(2)}$)
is unsuitable for perturbative approaches. This problem is overcome
by introducing the reaction matrix $G$ given by the solution of the
Bethe--Goldstone equation
\begin{equation}
    G=V+V\frac{Q}{\omega - H_0}G,
\end{equation}
where $\omega$ is the unperturbed energy of the interacting nucleons,
and $H_0$ is the unperturbed hamiltonian. 
The operator $Q$, commonly referred to
as the Pauli operator, is a projection operator which prevents the
interacting nucleons from scattering into states occupied by other nucleons.
In this work we solve the Bethe--Goldstone equation for five starting
energies $\Omega$, by way of the so--called double--partitioning scheme
discussed in e.g.,  Ref.\ \cite{hko95}. 
The $G$-matrix is the sum over all
ladder type of diagrams. This sum is meant to renormalize
the repulive short--range part of the interaction. The physical interpretation
is that the particles must interact with each other an infinite number
of times in order to produce a finite interaction. 
To construct the Pauli operator which defines $G$, one has to take 
into account that neutrons and protons have different closed shell
cores, $N=82$ and $Z=50$, respectively.
This means that neutrons in the $2s1d0g_{7/2}0h_{11/2}$ shell
are holes, while protons in the $2s1d0g_{7/2}0h_{11/2}$ shell
are particles. For protons the  Pauli operator must be constructed so as to
prevent scattering into intermediate states with a single
proton in any of the
states defined by the orbitals from the $0s$ shell up to the
$0g_{9/2}$ orbital. For a two--particle state with protons only, one has 
also to avoid scattering into states with two protons in the $2s1d0g$ 
($0g_{9/2}$ excluded) and $2p1f0h$ shells. For neutrons one must
prevent scattering into intermediate states with a single neutron
in the orbitals from the $0s$ shell up to the
$0h_{11/2}$ orbital. In addition, in the case of a two-particle
state with neutrons only, one must prevent scattering into states
with two neutrons in the $0h_{9/2}0i_{13/2}1f2p$ and 
$3s2d1g0i_{11/2}0j_{15/2}$ shells. 
If we have a proton--neutron two--particle state we must in addition prevent
scattering into two--body states where a proton is in the 
the $2s1d0g$ ($0g_{9/2}$ excluded)
and $2p1f0h$ shells and a neutron is in the 
$0h_{9/2}0i_{13/2}1f2p$ and $3s2d1g0i_{11/2}0j_{15/2}$ shells.

A harmonic--oscillator basis was chosen for the
single-particle
wave functions, with an oscillator energy $\hbar\Omega$ given
by
$\hbar\Omega = 45A^{-1/3} - 25A^{-2/3}=7.87 $ MeV,  $A=132$ being the mass
number.

Finally, we briefly sketch how to calculate an effective 
two-body interaction for the chosen model space
in terms of the $G$--matrix.  Since the $G$--matrix represents just
the summmation to all orders of ladder diagrams with particle-particle
diagrams, there are obviously other terms which need to be included
in an effective interaction. Long--range effects represented by 
core--polarization terms are also needed.
The first step then is to define the so--called $\hat{Q}$-box given by
\begin{equation}
   P\hat{Q}P=PGP +
   P\left(G\frac{Q}{\omega-H_{0}}G\\ + G
   \frac{Q}{\omega-H_{0}}G \frac{Q}{\omega-H_{0}}G +\dots\right)P.
   \label{eq:qbox}
\end{equation}
The $\hat{Q}$--box is made up of non--folded diagrams which are irreducible
and valence linked. A diagram is said to be irreducible if between each pair
of vertices there is at least one hole state or a particle state outside
the model space. In a valence--linked diagram the interactions are linked
(via fermion lines) to at least one valence line. Note that a valence--linked
diagram can be either connected (consisting of a single piece) or
disconnected. In the final expansion including folded diagrams as well, the
disconnected diagrams are found to cancel out \cite{ko90}.
This corresponds to the cancellation of unlinked diagrams
of the Goldstone expansion \cite{ko90}.
These definitions are discussed in Refs.\ \cite{hko95,ko90}.
We can then obtain an effective interaction
$H_{\mathrm{eff}}=\tilde{H}_0+V_{\mathrm{eff}}^{(2)}$ in terms of the $\hat{Q}$--box,
with \cite{hko95,ko90}
\begin{equation}
    V_{\mathrm{eff}}^{(2)}(n)=\hat{Q}+{\displaystyle\sum_{m=1}^{\infty}}
    \frac{1}{m!}\frac{d^m\hat{Q}}{d\omega^m}\left\{
    V_{\mathrm{eff}}^{(2)}(n-1)\right\}^m,
    \label{eq:fd}
\end{equation}
where $(n)$ and $(n-1)$ refer to the effective interaction after
$n$ and $n-1$ iterations. The zeroth iteration is represented by just the 
$\hat{Q}$--box.
Observe also that the
effective interaction $V_{\mathrm{eff}}^{(2)}(n)$
is evaluated at a given model space energy
$\omega$, as is the case for the $G$--matrix as well. Here we choose
$\omega =-20$ MeV.
Moreover, although $\hat{Q}$ and its derivatives contain disconnected
diagrams, such diagrams cancel exactly in each order \cite{ko90}, thus
yielding a fully connected expansion in e.g.\ Eq.\ (\ref{eq:fd}).
Less than $10$ iterations were needed in order to obtain a numerically
stable result. All non--folded diagrams through 
third--order in the interaction $G$ are included.
For further details, see Ref.\ \cite{hko95}.

The effective two--particle interaction can in turn be used in shell model
calculations.
The shell model problem requires the solution of a real symmetric
$n \times n$ matrix eigenvalue equation
\begin{equation}
       \tilde{H}\left | \Psi_k\right\rangle  = 
       E_k \left | \Psi_k\right\rangle ,
       \label{eq:shell_model}
\end{equation}
with $k = 1,\ldots, K$. 
At present our basic approach to 
finding solutions to Eq.\ (\ref{eq:shell_model})
is the Lanczos algorithm; an iterative method which gives the solution of
the lowest eigenstates. This method was 
already applied to nuclear physics problems by Whitehead {\sl et al.} 
in 1977. The technique is described in detail in Ref.\ \cite{whit77}, 
see also \cite{ehho95}. 

The eigenstates of Eq.\ (\ref{eq:shell_model}) are
written as linear combinations of Slater determinants in the $m$--scheme,
distributing the $N$ particles(holes) in all possible
ways through the single particle $m$--scheme orbitals of the 
model space,  
$2s_{1/2}$, $1d_{5/2}$, $1d_{3/2}$, $0g_{7/2}$ and $0h_{11/2}$.
As seen in Table \ref{tab:table1}, the dimensionality
$n$ of the eigenvalue matrix $\tilde{H}$ is increasing
with increasing number of valence holes, and
for the Sn isotopes of interest it is up to $n \approx 2 \times 10^{7}$.
\begin{table}[htbp]
\begin{center}
\caption{Number of basis states for the shell model calculation of the $N=82$
isotopes, with $1d_{5/2}$, $0g_{7/2}$, $1d_{3/2}$, $2s_{1/2}$ and $0h_{11/2}$
single particle orbitals.}
\begin{tabular}{lrcrlr}-
\\\hline
System & Dimension & System & Dimension & System & Dimension \\
\hline
$^{130}$Sn & 36        & $^{125}$Sn & 108 297   & $^{120}$Sn & 6 210 638 \\
$^{129}$Sn  & 245       & $^{124}$Sn & 323 682   & $^{119}$Sn & 9 397 335 \\ 
$^{128}$Sn & 1 504     & $^{123}$Sn & 828 422   & $^{118}$Sn & 12 655 280 \\
$^{127}$Sn & 7 451     & $^{122}$Sn & 1 853 256 & $^{117}$Sn & 15 064 787 \\
$^{126}$Sn & 31 124    & $^{121}$Sn & 3 609 550 & $^{116}$Sn & 16 010 204 \\
\hline
\end{tabular}
\label{tab:table1}
\end{center}
\end{table}

\section{Results and discussions}

In this section we present and discuss the shell model results for as many 
as 12 valence nucleons. The intention is to gain insight and draw conclusions 
regarding the effective interaction on basis of a spectroscopic analysis. The
effective interaction is derived for a core with $N \neq Z$. 

All experimental information in the present analysis is taken from the data 
base of the National Nuclear Data Center at Brookhaven \cite{nndc}.

\subsection{Even tin isotopes}

We will here discuss the even Sn isotopes as a whole, and focus on the 
general spectroscopic trends. The results of the SM calculation are displayed
in Tables \ref{tab:sn130} -- \ref{tab:sn120}. The energy eigenvalues are
sorted according to the angular momentum assignment. 

\begin{table}[htbp]
\begin{center}
\begin{tabular}{cccc|cccc}
\hline
\multicolumn{4}{c}{ $^{130}$Sn} & \multicolumn{4}{c}{ $^{128}$Sn} \\ 
{$J^{\pi}_i$} & {Exp} & {$J^{\pi}_i$} & {Theory} & 
{$J^{\pi}_i$} & {Exp} & {$J^{\pi}_i$} & {Theory} \\
\hline 
$0^{+}$ & 0.00 & $0^{+}_{1}$ & 0.00 &
  $0^{+}$ & 0.00 & $0^{+}_{1}$ & 0.00 \\
        &      & $0^{+}_{2}$ & 2.11 &
          &      & $0^{+}_{2}$ & 2.33 \\
        &      & $0^{+}_{3}$ & 2.38 &
          &      & $0^{+}_{3}$ & 2.52 \\
$(2^{+})$ & 1.22 & $2^{+}_{1}$ & 1.46 &
  $(2^{+})$  & 1.17 & $2^{+}_{1}$ & 1.28 \\
$(2^{+})$ & 2.03 & $2^{+}_{2}$ & 2.17 &
  $(1,2)^{+}$ & 2.10 & $2^{+}_{2}$ & 2.14 \\
          &      & $2^{+}_{3}$ & 2.46 &
  $1,2^{+}$   & 2.26 & $2^{+}_{3}$ & 2.52 \\
$(4^{+})$ & 2.00 & $4^{+}_{1}$ & 2.39 &
  $(4^{+})$ & 2.00 & $4^{+}_{1}$ & 2.18 \\
$(4^{+})$ & 3.42 & $4^{+}_{2}$ & 3.23 &
            &      & $4^{+}_{2}$ & 2.84 \\
$(6^{+})$ & 2.26 & $6^{+}_{1}$ & 2.64 &
  $(6,7^{-})$ & 2.38 & $6^{+}_{1}$ & 2.53 \\
$(8^{+})$ & 2.34 & $8^{+}_{1}$ & 2.72 &
  $(7^{+},8,9)$ & 2.41 & $8^{+}_{1}$ & 2.66 \\
$(10^{+})$ & 2.44 & $10^{+}_{1}$ & 2.80 &
                &     & $10^{+}_{1}$ & 2.80 \\
%
$(3^{-},4^{+})$ & 2.49 & $3^{-}_{1}$ & 3.44 &
                &      & $3^{-}_{1}$ & 3.11 \\
$(3^{-},4^{+})$ & 4.22 & $3^{-}_{2}$ & 4.67 &
                &      & $3^{-}_{2}$ & 3.25 \\
$(5^{-})$ & 2.09 & $5^{-}_{1}$ & 2.19 &
  $(5^{-})$ & 2.12 & $5^{-}_{1}$ & 2.27 \\
$(7^{-})$ & 1.95 & $7^{-}_{1}$ & 2.03 &
  $(7^{-})$ & 2.09 & $7^{-}_{1}$ & 2.28 \\
&&&\\ \hline
\end{tabular}
\caption{Low--lying states for  $^{130}$Sn and $^{128}$Sn. Energies in MeV.}
\label{tab:sn130}
\end{center}
\end{table}

\begin{table}[htbp]
\begin{center}
\begin{tabular}{cccc|cccc}
\hline
\multicolumn{4}{c}{ $^{126}$Sn} & \multicolumn{4}{c}{ $^{124}$Sn} \\ 
{$J^{\pi}_i$} & {Exp} & {$J^{\pi}_i$} & {Theory} & 
{$J^{\pi}_i$} & {Exp} & {$J^{\pi}_i$} & {Theory} \\
\hline 
$0^{+}$ & 0.00 & $0^{+}_{1}$ & 0.00 &
  $0^{+}$   & 0.00 & $0^{+}_{1}$ & 0.00 \\
            &      & $0^{+}_{2}$ & 2.20 &
$(0^{+})$ & 2.13 & $0^{+}_{2}$ & 2.30 \\
            &      & $0^{+}_{3}$ & 2.75 &
$(0^{+})$ & 2.69 & $0^{+}_{3}$ & 2.84 \\
$2^{+}$ & 1.14 & $2^{+}_{1}$ & 1.21 &
  $2^{+}  $ & 1.13 & $2^{+}_{1}$ & 1.17 \\
$2^{+}$ & 2.11 & $2^{+}_{2}$ & 2.17 &
   $(2^{+})$ & 2.12 & $2^{+}_{2}$ & 2.16 \\
$2^{+}$ & 2.37 & $2^{+}_{3}$ & 2.60 &
  $(2^{+})$ & 2.43 & $2^{+}_{3}$ & 2.73 \\
$4^{+}$     & 2.05 & $4^{+}_{1}$ & 2.21 &
  $4^{+}  $ & 2.10 & $4^{+}_{1}$ & 2.26 \\
$2,3,4^{+}$ & 2.71 & $4^{+}_{2}$ & 2.64 &
  $(2,3,4)$ & 2.60 & $4^{+}_{2}$ & 2.53 \\
$4^{+}$     & 3.42 & $4^{+}_{2}$ & 3.09 &
  $(4^{+})$ & 2.70 & $4^{+}_{3}$ & 3.03 \\
            &      & $6^{+}_{1}$ & 2.61 &
            &      & $6^{+}_{1}$ & 2.70 \\
            &      & $8^{+}_{1}$ & 2.74 &
  $(8^{+})$ & 2.45 & $8^{+}_{1}$ & 2.80 \\
            &     & $10^{+}_{1}$ & 2.77 &
            &     & $10^{+}_{1}$ & 2.85 \\
$3^{-}$ & 2.72 & $3^{-}_{1}$ & 3.04 &
  $3^{-}$ & 2.61 & $3^{-}_{1}$ & 2.97 \\
$3^{-}$ & 4.77 & $3^{-}_{2}$ & 3.21 &
  $(3^{-})$ & 3.01 & $3^{-}_{2}$ & 3.35 \\
$5^{-}$ & 2.16 & $5^{-}_{1}$ & 2.36 &
 $5^{-}$ & 2.21 & $5^{-}_{1}$ & 2.46 \\
$5^{-}$ & 2.89 & $5^{-}_{2}$ & 2.76 &
         &      & $5^{-}_{2}$ & 2.84 \\
$7^{-}$ & 2.22 & $7^{-}_{1}$ & 2.48 &
 $7^{-}$ & 2.33 & $7^{-}_{1}$ & 2.63 \\
&&&\\ \hline
\end{tabular}
\caption{Low--lying states for  $^{126}$Sn and $^{124}$S. Energies in MeV.}
\label{tab:sn126}
\end{center}
\end{table}

\begin{table}[htbp]
\begin{center}
\begin{tabular}{cccc|cccc}
\hline
\multicolumn{4}{c}{ $^{122}$Sn} & \multicolumn{4}{c}{ $^{120}$Sn}\\ 
{$J^{\pi}_i$} & {Exp} & {$J^{\pi}_i$} & {Theory} & 
{$J^{\pi}_i$} & {Exp} & {$J^{\pi}_i$} & {Theory} \\
\hline 
$0^{+}$ & 0.00 & $0^{+}_{1}$ & 0.00 &
  $0^{+}$ & 0.00 & $0^{+}_{1}$ & 0.00 \\
$0^{+}$ & 2.09 & $0^{+}_{2}$ & 2.41 &
  $0^{+}$ & 1.88 & $0^{+}_{2}$ & 2.51 \\
($0^{+})$ & 2.67 & $0^{+}_{3}$ & 2.80 & 
  $0^{+}$ & 2.16 & $0^{+}_{3}$ & 2.66 \\
$2^{+}$ & 1.14 & $2^{+}_{1}$ & 1.15 &
  $2^{+}$ & 1.17 & $2^{+}_{1}$ & 1.14 \\
$2^{+}$ & 2.15 & $2^{+}_{2}$ & 2.15 & 
  $2^{+}$ & 2.10 & $2^{+}_{2}$ & 2.13 \\
$4^{+}$ & 2.14 & $4^{+}_{1}$ & 2.30 &
  $4^{+}$ & 2.19 & $4^{+}_{1}$ & 2.30 \\
$4^{+}$ & 2.33 & $4^{+}_{2}$ & 2.51 &
  $(4^{+})$ & 2.47 & $4^{+}_{2}$ & 2.54 \\
$6^{+}$ & 2.56 & $6^{+}_{1}$ & 2.78 &
          &      & $6^{+}_{1}$ & 2.86 \\
$8^{+}$ & 2.69 & $8^{+}_{1}$ & 2.88 &
  $(8^{+})$ & 2.69 & $8^{+}_{1}$ &      \\
$10^{+}$ & 2.78 & $10^{+}_{1}$ & 2.95 &
          &     & $10^{+}_{1}$ &      \\
$3^{-}$ & 2.49 & $3^{-}_{1}$ & 2.90 &
  $3^{-}$ & 2.40 & $3^{-}_{1}$ & 2.86 \\
$3^{-}$ & 3.36 & $3^{-}_{2}$ & 3.47  &
  $3^{-}$ & 3.47 & $3^{-}_{2}$ & \\
$5^{-}$ & 2.25 & $5^{-}_{1}$ & 2.55 &
  $5^{-}$ & 2.28 & $5^{-}_{1}$ & 2.63 \\
$5^{-}$ & 2.75 & $5^{-}_{2}$ & 2.96 &
  $(5^{-})$ & 2.55 & $5^{-}_{2}$ & 3.10 \\
%
$7^{-}$ & 2.41 & $7^{-}_{1}$ & 2.74 &
  $7^{-}$ & 2.48 & $7^{-}_{1}$ & 2.85 \\
&&&\\ \hline
\end{tabular}
\caption{Low--lying states for  $^{122}$Sn and $^{120}$Sn. Energies in MeV.}
\label{tab:sn120}
\end{center}
\end{table}

The stable energy spacing between the $0^{+}$ ground state and the first
excited $2^{+}$ state, which is so characteristic for the Sn isotopes, is well
reproduced in our shell model calculation.
The yrast states, $4^{+}_{1}$, $6^{+}_{1}$, $8^{+}_{1}$ and $10^{+}_{1}$ 
tend to have slightly too high excitation energies. The calculated energies
are 0.1 -- 0.4 MeV higher than the experimental ones.
Non--yrast states like $2^+_2$ and $4^+_2$ are again in beautiful agreement 
with experiment. 

The agreement between the calculated and the experimental $5^-_1$ and $7^-_1$ 
states in $^{130}$Sn is very satisfactory. Towards the middle of the shell
the deviation between theory and experiment becomes larger. The
experimental excitation energies of the $5^-_1$ and $7^-_1$ states increase 
slightly as approaching the middle of the shell. Our calculated energy 
levels do however increase too much and come out about 0.3 MeV too high in 
$^{120}$Sn. 

The $3^-$ data is not well established in $^{130}$Sn and 
$^{128}$Sn, but where a comparison is possible the calculated $3^{-}_{1}$ 
states come out 0.3 -- 0.4 MeV higher than their experimental counterparts.
The calculated $3^{-}_{2}$ states are also too high in excitation energy.
It may be of interest to note that in the neighbouring N=82 isotones 
low--lying $3^{-}$ states are observed which cannot be reproduced by the 
shell model \cite{holt97} and thus presumably are collective states. 
For the tin isotopes we obtain  two $3^{-}$ states, which are somewhat 
too strongly excited, but it is likely that both states are of shell model
nature.

Excited $0^{+}$ states are at present only observed in the tin isotopes
with $N<76$. Towards the middle of the N=50-82 shell, the $0^{+}_{2}$ state 
comes gradually lower in energy. It is worth mentioning that at midshell, 
in $^{116}$Sn, $0^{+}_{2}$ is the first excited state.
The agreement between the calculated and the observed $0^{+}_{2}$ state in 
$^{124}$Sn is rather good. The deviation between our 
calculated $0^{+}_{2}$ state and the observed first excited $0^{+}$ state 
in $^{122}$Sn and $^{120}$Sn is 0.32 MeV and 0.63 MeV, respectively.
We do not manage to reproduce the trend with decreasing excitation energy of
the first excited $0^{+}$ state as approaching the middle of the shell.

Close to midshell low--lying $0^+$ intruder states have been reported 
that are supposed to be predominantly proton core excitations. The first 
evidence for low--lying proton excitations across the $Z=50$ gap was found 
in $^{108-118}$Sn by Fielding {\em et al.} \cite{fiel77}. Later, collective 
band structures have been identified by Bron {\em et al.} \cite{bron79},
bands which are assumed to be based on deformed proton configurations.
Such states are however beyond the scope of the present model.

\subsubsection{Generalized seniority for even isotopes}

We will here make a comparison of the SM with the generalized
seniority model \cite{talm77}. The generalized seniority scheme is an 
extension of the seniority scheme, i.e. from involving only one single 
$j$--orbital, the model is generalized to involve a group of $j$--orbitals 
within a major shell. The generalized seniority scheme is a more simple 
model than the shell model since a rather limited number of configurations 
with a strictly defined structure are included, thus allowing a more
direct physical interpretation. States with seniority $v=0$ are by 
definition states where all particles are coupled in pairs. Seniority 
$v=2$ states have one pair broken, seniority $v=4$ states have two pairs 
broken, etc. The generalized seniority scheme is suitable for describing 
semi--magic nuclei where pairing plays an important role. The pairing picture 
and the generalized seniority scheme have been important for the description 
and understanding of the tin isotopes. A typical feature of the seniority 
scheme is that the spacing of energy levels are independent of the number of 
valence particles. For the tin isotopes, not only the spacing between the 
ground state and the $2^{+}_{1}$ state, but also the spacing beween the 
ground state and the $4^{+}_{1}$ and $6^{+}_{1}$ states is fairly constant 
throughout the whole sequence of isotopes. In fundamental works on generalized 
seniority by Talmi \cite{talm77}, the tin isotopes have been used as one of 
the major test cases. It is also worth mentioning the classical work on 
pairing by Kisslinger and Sorensen \cite{kiss60}. 

If we by closer investigation and comparison of the SM wave function and the
seniority states find that the most important components are accounted for  
by the seniority scheme, we can benefit from this and reduce the SM basis. 
This would be particularly useful when we want to do calculations on systems 
with a large number of valence particles.

The operator for creating a generalized seniority $(v=0)$ pair is
\begin{equation}
    S^{\dagger}= \sum_{j} 
    \frac{1}{\sqrt{2j+1}}C_{j}\sum_{m \geq 0} (-1)^{j-m} b^{\dagger}_{jm}
    b_{j-m}^{\dagger},
    \label{eq:s_dagger}
\end{equation}
where  $b^{\dagger}_{jm}$ is the creation operator for holes.
The generalized senitority $(v=2)$ operator for creating a broken pair 
is given by
\begin{equation}
     D_{J,M}^{\dagger}=
     \sum_{j \leq j'} (1+\delta_{j,j'})^{-1/2}
     \beta_{j,j'} \langle j m j' m'
     \left | JM \right \rangle b^{\dagger}_{jm}b^{\dagger}_{j'm'}.
\end{equation}
The coefficients $C_{j}$ and $\beta_{jj'}$ are obtained from the $^{130}$Sn
ground state and the excited states, respectively.

We calculate the squared overlaps between the constructed generalized seniority
states and our shell model states
\begin{equation}
\begin{array}{lccccl}
(v=0) & & & & &
|\langle ^{A}{\rm Sn(SM)} ;
0^{+}|(S^{\dagger})^{\frac{n}{2}}| \tilde{0} \rangle |^{2}, \\
(v=2) & & & & &
|\langle ^{A}{\rm Sn(SM)} ;
J_{i}|D^{\dagger}_{JM}(S^{\dagger})^{\frac{n}{2}-1}| \tilde{0} \rangle |^{2}. 
\end{array}
\end{equation}
The vacuum state $|\tilde{0} \rangle $ is the $^{132}$Sn--core and $n$ is
the number of valence particles. These quantities tell to what extent the 
shell model states satisfy the pairing picture, or in other words, how well 
is generalized seniority conserved as a quantum number.

The squared overlaps are tabulated in Table~\ref{tab:seniority}, and vary
generally from 0.95 to 0.75. As the number of valence particles increases the 
squared overlaps are gradually decreasing. The overlaps involving the $4^{+}$
states show a fragmentation. In $^{128}$Sn, the $4^{+}_{1}$ (SM)
state is mainly a seniority $v=2$ state. As approaching the middle of the 
shell, the next state, $4^{+}_{2}$, takes more and more over the 
structure of a seniority $v=2$ state. The fragmentation of seniority over these
two states can be understood from the fact that they are rather close in 
energy and therefore may have mixed structure.

\begin{table}[htbp]
\begin{center}
\begin{tabular}{cccccc}
\hline
 & A=128 & A=126 & A=124 & A=122 & A=120 \\ 
\hline
$0^{+}_{1}$ & 0.96 & 0.92 & 0.87 & 0.83 & 0.79 \\ 
$2^{+}_{1}$ & 0.92 & 0.89 & 0.84 & 0.79 & 0.74 \\ 
$4^{+}_{1}$ & 0.73 & 0.66 & 0.44 & 0.13 & 0.00 \\ 
$4^{+}_{2}$ & 0.13 & 0.18 & 0.39 & 0.66 & 0.74 \\
$6^{+}_{1}$ & 0.81 & 0.85 & 0.83 & 0.79 & 0.64 \\
\hline
\end{tabular}
\caption{ Seniority $v=0$ overlap 
         $|\langle ^{A}Sn;0^{+}|(S^{\dagger})^{\frac{n}{2}}| 
         \tilde{0} \rangle |^{2}$ and the seniority $v=2$ 
         overlaps $|\langle ^{A}Sn ;J_{f}|
         D^{\dagger}_{JM}(S^{\dagger})^{\frac{n}{2} - 1}| 
         \tilde{0} \rangle |^{2}$ for the lowest--lying eigenstates 
         of $^{128-120}$Sn.}
\label{tab:seniority}
\end{center}
\end{table}

\subsubsection{Effective charges and E2 transitions}

In the literature there is some discussion concerning the effective charges 
to be used in B(E2) calculations. Fogelberg and Blomqvist have used an 
effective charge $e_{\rm eff} = 1.0 e$ in their study of single--hole states 
in $^{131}$Sn \cite{foge84}. In a work on $\nu h_{11/2}$ subshell filling, 
Broda {\em et al.} \cite{brod92} are assigning an effective neutron charge, 
$e_{\rm eff} = 0.88(4)e$, to the $h_{11/2}$ holes. State dependent effective 
charges can be derived along the same lines as the effective interaction. 
Preliminary calculations of the effective charges for this mass region  
by one of the authors, Ref.\ \cite{hjor97}, have given values between 
$0.7e$ and $0.9e$ for the different valence particle states. 
However, in this work the state dependence has not been incorporated, but an 
average value $e_{\rm eff} = 0.8e$ has been used.

In the vicinity of the $^{132}$Sn--core E2 transition data are rather 
scarce. Towards the middle of the shell, and more stable isotopes, data
are available. The agreement between our calculated B(E2) values and 
experimental data is overall very good, see Tables \ref{tab:rateof} and 
\ref{tab:rateofe2}. In $^{122}$Sn and $^{120}$Sn
several E2 transition rates are measured. The experimental 
$B(E2;4^+_1 \rightarrow 2^+_1)$ and $B(E2;2^+_1 \rightarrow 0^+_1)$ values
are measured to be 10 -- 20 W.u., which indicates that the states are of 
collective nature. We reproduce all these transitions exceptionally well. For 
the two nuclei there is also experimental information on the transition 
$0^+_2 \rightarrow 2_1^+$. In $^{122}$Sn an upper limit,
$B(E2;2^+_1 \rightarrow 0^+_1) < 6.1$ W.u. is given, and in $^{120}$S
the transition is reported to be 18.6 W.u., i.e. strongly collective. 
The transition rates are very sensitive to the structure of the 
wave functions and thus can give us valuable information. We fail in 
reproducing these E2 transitions by several orders of magnitude, and from 
that we can conclude that our SM $0^+_2$ wave functions do not have the 
correct structure. As mentioned in the previous subsection the SM does also 
have problems in reproducing the correct energy of these $0^+_2$ states. 
The calculated energies are too high compared to their experimental 
counterparts.

The measured $B(E2;7^-_1 \rightarrow 5^-_1)$ transition rates are all small,
0.29 W.u. in $^{126}$Sn and reduced to 0.004 W.u. in $^{120}$Sn. While the 
experimental transition rates decrease with the number of valence particles,
our calculated transition rates increase from 0.085 W.u. in $^{126}$Sn to
2.90 W.u. in $^{120}$Sn. This may indicate some short-comings of our negative
parity wave functions. We believe that is associated with the interplay 
between the $0h_{11/2}$ orbital and the other orbitals. Even small 
corrections may change such weak transitions.

Broda {\em et al.} \cite{brod92} argue that an effective charge twice as 
large as the value for $^{130}$Sn is needed in $^{116}$Sn. This enhancement
of the $e_{\rm eff}$ in the middle of the N = 50 -- 82 shell is interpreted 
as a consequence of additional configuration mixing, corresponding to 
polarization of the softer midshell core. Our E2 results for 
$^{122,120}$Sn do however not indicate the need for any larger effective 
charge as approaching the middle of the shell.

\begin{table}[htbp]
\begin{center}
\begin{tabular}{ccccccc}
\hline
 & \multicolumn{2}{c}{ $^{130}$Sn} &  \multicolumn{2}{c}{ $^{128}$Sn} &
\multicolumn{2}{c}{ $^{126}$Sn} \\

Transition& SM & Exp. & SM & Exp & SM & Exp \\ \hline
$2_{1}^{+}$ $\rightarrow$ $0_{1}^{+}$  & 2.07 &          & 4.33 & & 6.57 & \\ 
$0_{2}^{+}$ $\rightarrow$ $2_{1}^{+}$  & 0.96 &          & 0.056 & & 0.17 & \\
$4_{1}^{+}$ $\rightarrow$ $2_{1}^{+}$  & 1.45 &          & 4.13 & & 6.03 & \\  
$5_{1}^{-}$ $\rightarrow$ $7_{1}^{-}$  & 1.18 & 1.4 (2)  & & & &          \\
$7_{1}^{-}$ $\rightarrow$ $5_{1}^{-}$  &      &          & & & 0.085 & 0.29 (5) \\
$10_{1}^{+}$ $\rightarrow$ $8_{1}^{+}$ & 0.35 & 0.38 (4) & & & &          \\
&&&&&&\\
\hline
\end{tabular}
\caption{E2--transitions for $^{130,128,126}$Sn. Units in W.u.}
\label{tab:rateof}
\end{center}
\end{table}

\begin{table}[htbp]
\begin{center}
\begin{tabular}{ccccccc}
\hline
 & \multicolumn{2}{c}{ $^{124}$Sn} &  \multicolumn{2}{c}{ $^{122}$Sn} &
\multicolumn{2}{c}{ $^{120}$Sn} \\
Transition& SM & Exp & SM & Exp & SM & Exp \\ 
\hline \\
$2_{1}^{+}$ $\rightarrow$ $0_{1}^{+}$ & 8.70 & 9.0 (2)  & 10.6 & 
                                          10.7 (6)   & 12.0  & 16.9 (3) \\ 
$0_{2}^{+}$ $\rightarrow$ $2_{1}^{+}$ & 0.38 &          & 0.046 & 
                                            $<$ 6.1  & 0.007 & 18.6 (25) \\ 
$7_{1}^{-}$ $\rightarrow$ $5_{1}^{-}$ & 0.64 & 0.11 (2) & 1.80 & 
                                         0.0146 (21) & 2.90  & 0.0040 (2) \\
$4_{1}^{+}$ $\rightarrow$ $2_{1}^{+}$ & 9.42 &          & 14.6 & 
                                         10.0 (14)   & 18.2  & 14.8 (21) \\
&&&&\\
\hline
\end{tabular}
\caption{E2--transitions for $^{124,122,120}$Sn. Units in W.u.}
\label{tab:rateofe2}
\end{center}
\end{table}

\subsection{Odd isotopes}

\begin{figure}[htbp]
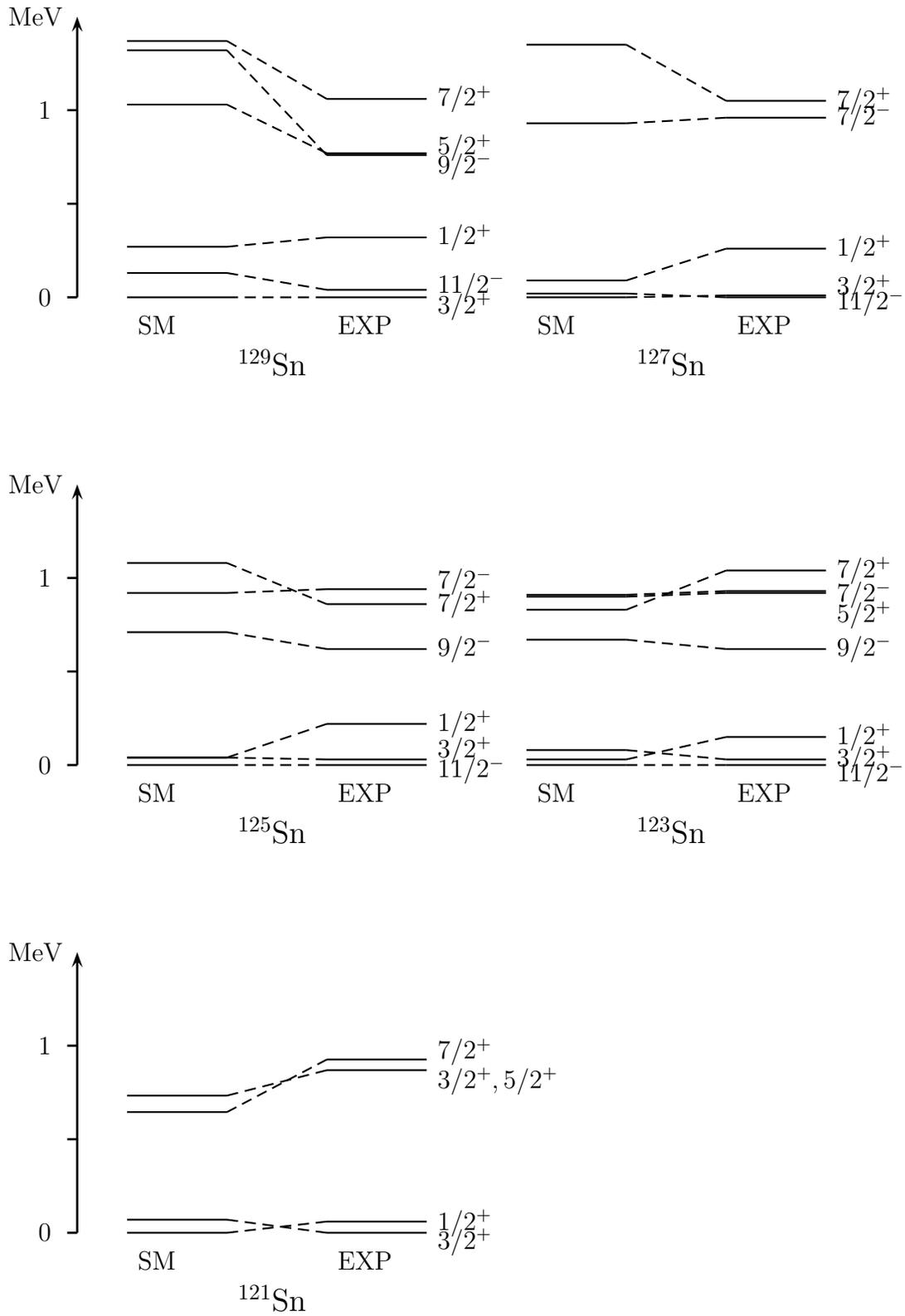

 \setlength{\unitlength}{1cm}
 \begin{center}
 \setlength{\unitlength}{1cm}
 \thicklines

 \Cartesian(1.6cm,3cm)
 \pspicture(0,0)(10,7)
\psline[linewidth=1pt]{->}(0.5,5.5)(0.5,7)

\psline[linewidth=1pt]{-}(0.4,5.5)(0.5,5.5)
\psline[linewidth=1pt]{-}(0.4,6)(0.5,6)
\psline[linewidth=1pt]{-}(0.4,6.5)(0.5,6.5)

\uput[0](0,5.5){0}
\uput[0](0,6.5){1}

\uput[0](-0.3,7){MeV}

\psline{-}(1,5.5)(2,5.5) 
\psline[linestyle=dashed]{-}(2,5.5)(3,5.5)
\psline{-}(3,5.5)(4,5.5)

\psline{-}(1,5.63)(2,5.63) 
\psline[linestyle=dashed]{-}(2,5.63)(3,5.54)
\psline{-}(3,5.54)(4,5.54)

\psline{-}(1,5.77)(2,5.77) 
\psline[linestyle=dashed]{-}(2,5.77)(3,5.82)
\psline{-}(3,5.82)(4,5.82)

\psline{-}(1,6.53)(2,6.53) 
\psline[linestyle=dashed]{-}(2,6.53)(3,6.27)
\psline{-}(3,6.27)(4,6.27)

\psline{-}(1,6.87)(2,6.87) 
\psline[linestyle=dashed]{-}(2,6.87)(3,6.56)
\psline{-}(3,6.56)(4,6.56)

\psline{-}(1,6.82)(2,6.82) 
\psline[linestyle=dashed]{-}(2,6.82)(3,6.26)
\psline{-}(3,6.26)(4,6.26)

\uput[0](4,5.45){$3/2^{+}$}
\uput[0](4,5.56){$11/2^{-}$}
\uput[0](4,5.82){$1/2^{+}$}
\uput[0](4,6.30){$5/2^{+}$}
\uput[0](4,6.56){$7/2^{+}$}
\uput[0](4,6.20){$9/2^{-}$}

\psline{-}(5,5.5)(6,5.5) 
\psline[linestyle=dashed]{-}(6,5.5)(7,5.51)
\psline{-}(7,5.51)(8,5.51)

\psline{-}(5,5.52)(6,5.52) 
\psline[linestyle=dashed]{-}(6,5.52)(7,5.5)
\psline{-}(7,5.5)(8,5.5)

\psline{-}(5,5.59)(6,5.59) 
\psline[linestyle=dashed]{-}(6,5.59)(7,5.76)
\psline{-}(7,5.76)(8,5.76)

\psline{-}(5,6.85)(6,6.85)
\psline[linestyle=dashed]{-}(6,6.85)(7,6.55)
\psline{-}(7,6.55)(8,6.55)

\psline{-}(5,6.43)(6,6.43)
\psline[linestyle=dashed]{-}(6,6.43)(7,6.46)
\psline{-}(7,6.46)(8,6.46)

\uput[0](8,5.47){$11/2^{-}$}
\uput[0](8,5.55){$3/2^{+}$}
\uput[0](8,5.76){$1/2^{+}$}
\uput[0](8,6.55){$7/2^{+}$}
\uput[0](8,6.46){$7/2^{-}$}

\uput[0](1,5.35){SM}
\uput[0](3,5.35){EXP}
\uput[0](2,5.15){\large $^{129}$Sn}

\uput[0](5,5.35){SM}
\uput[0](7,5.35){EXP}
\uput[0](6,5.15){\large $^{127}$Sn}
\psline[linewidth=1pt]{->}(0.5,3)(0.5,4.5)

\psline[linewidth=1pt]{-}(0.4,3)(0.5,3)
\psline[linewidth=1pt]{-}(0.4,3.5)(0.5,3.5)
\psline[linewidth=1pt]{-}(0.4,4)(0.5,4)

\uput[0](0,3){0}
\uput[0](0,4){1}
\uput[0](-0.3,4.5){MeV}

\psline{-}(1,3)(2,3) 
\psline[linestyle=dashed]{-}(2,3)(3,3)
\psline{-}(3,3)(4,3)

\psline{-}(1,3.04)(2,3.04) 
\psline[linestyle=dashed]{-}(2,3.04)(3,3.03)
\psline{-}(3,3.03)(4,3.03)

\psline{-}(1,3.04)(2,3.04) 
\psline[linestyle=dashed]{-}(2,3.04)(3,3.22)
\psline{-}(3,3.22)(4,3.22)

\psline{-}(1,4.08)(2,4.08) 
\psline[linestyle=dashed]{-}(2,4.08)(3,3.86)
\psline{-}(3,3.86)(4,3.86)

\psline{-}(1,3.71)(2,3.71) 
\psline[linestyle=dashed]{-}(2,3.71)(3,3.62)
\psline{-}(3,3.62)(4,3.62)

\psline{-}(1,3.92)(2,3.92) 
\psline[linestyle=dashed]{-}(2,3.92)(3,3.94)
\psline{-}(3,3.94)(4,3.94)

\uput[0](4,2.97){$11/2^{-}$}
\uput[0](4,3.08){$3/2^{+}$}
\uput[0](4,3.22){$1/2^{+}$}
\uput[0](4,3.86){$7/2^{+}$}
\uput[0](4,3.62){$9/2^{-}$}
\uput[0](4,3.98){$7/2^{-}$}

\psline{-}(5,3)(6,3) 
\psline[linestyle=dashed]{-}(6,3)(7,3)
\psline{-}(7,3)(8,3)

\psline{-}(5,3.08)(6,3.08) 
\psline[linestyle=dashed]{-}(6,3.08)(7,3.03)
\psline{-}(7,3.03)(8,3.03)

\psline{-}(5,3.03)(6,3.03) 
\psline[linestyle=dashed]{-}(6,3.03)(7,3.15)
\psline{-}(7,3.15)(8,3.15)

\psline{-}(5,3.90)(6,3.90)
\psline[linestyle=dashed]{-}(6,3.90)(7,3.92)
\psline{-}(7,3.92)(8,3.92)

\psline{-}(5,3.83)(6,3.83)
\psline[linestyle=dashed]{-}(6,3.83)(7,4.04)
\psline{-}(7,4.04)(8,4.04)

\psline{-}(5,3.67)(6,3.67)
\psline[linestyle=dashed]{-}(6,3.67)(7,3.62)
\psline{-}(7,3.62)(8,3.62)

\psline{-}(5,3.91)(6,3.91)
\psline[linestyle=dashed]{-}(6,3.91)(7,3.93)
\psline{-}(7,3.93)(8,3.93)

\uput[0](8,2.95){$11/2^{-}$}
\uput[0](8,3.03){$3/2^{+}$}
\uput[0](8,3.15){$1/2^{+}$}
\uput[0](8,3.80){$5/2^{+}$}
\uput[0](8,4.04){$7/2^{+}$}
\uput[0](8,3.62){$9/2^{-}$}
\uput[0](8,3.91){$7/2^{-}$}

\uput[0](1,2.85){SM}
\uput[0](3,2.85){EXP}
\uput[0](2,2.65){\large $^{125}$Sn}

\uput[0](5,2.85){SM}
\uput[0](7,2.85){EXP}
\uput[0](6,2.65){\large $^{123}$Sn}
\psline[linewidth=1pt]{->}(0.5,0.5)(0.5,2)

\psline[linewidth=1pt]{-}(0.4,0.5)(0.5,0.5)
\psline[linewidth=1pt]{-}(0.4,1)(0.5,1)
\psline[linewidth=1pt]{-}(0.4,1.5)(0.5,1.5)

\uput[0](0,0.5){0}
\uput[0](0,1.5){1}
\uput[0](-0.3,2){MeV}


\psline{-}(1,0.5)(2,0.5) 
\psline[linestyle=dashed]{-}(2,0.5)(3,0.56)
\psline{-}(3,0.56)(4,0.56)

\psline{-}(1,0.57)(2,0.57) 
\psline[linestyle=dashed]{-}(2,0.57)(3,0.5)
\psline{-}(3,0.5)(4,0.5)

\psline{-}(1,1.145)(2,1.145) 
\psline[linestyle=dashed]{-}(2,1.145)(3,1.426)
\psline{-}(3,1.426)(4,1.426)

\psline{-}(1,1.233)(2,1.233) 
\psline[linestyle=dashed]{-}(2,1.233)(3,1.369)
\psline{-}(3,1.369)(4,1.369)



\uput[0](4,0.45){$3/2^{+}$}
\uput[0](4,0.55){$1/2^{+}$}
\uput[0](4,1.309){$3/2^{+},5/2^{+}$}
\uput[0](4,1.466){$7/2^{+}$}

\uput[0](1,0.35){SM}
\uput[0](3,0.35){EXP}
\uput[0](2,0.15){\large $^{121}$Sn}

\endpspicture
 \end{center}
\caption{\label{fig:energy-level-1}Low--spin Y--rast states for $^{129-121}$Sn.}
 \end{figure}

The results for the odd nuclei $^{129 - 121}$Sn are tabulated in Tables
\ref{tab:129sn} -- \ref{tab:121sn}. In addition, for each nucleus, some 
selected yrast states are displayed in Fig. \ref{fig:energy-level-1}.

\begin{table}[htbp]
\begin{center}
\begin{tabular}{cccc}
\hline 
\multicolumn{4}{c}{ $^{129}$Sn} \\ 
{$J^{\pi}_i$} &{Exp} &{$J^{\pi}_i$} &{Theory} \\
\hline 
$(3/2^{+})$         & 0.00 & $3/2^{+}_{1}$  & 0.00 \\
$(11/2^{-})$        & 0.04 & $11/2^{-}_{1}$ & 0.13 \\
$(1/2^{+})$         & 0.32 & $1/2^{+}_{1}$  & 0.27 \\
$(9/2^{-})$         & 0.76 & $5/2^{+}_{1}$  & 1.03 \\
$(5/2^{+})$         & 0.77 & $7/2^{-}_{1}$  & 1.05 \\
$(7/2^{-})$         & 1.04 & $3/2^{+}_{2}$  & 1.22 \\
$(7/2^{+})$         & 1.06 & $15/2^{-}_{1}$ & 1.28 \\
$(1/2^{+},3/2^{+})$ & 1.22 & $9/2^{-}_{1}$  & 1.32 \\
$(1/2^{+},3/2^{+})$ & 1.29 & $7/2^{+}_{1}$  & 1.37 \\
$(7/2^{+})$         & 1.87 & $1/2^{+}_{2}$  & 1.43 \\
$(7/2^{+})$         & 2.12 & $3/2^{+}_{3}$  & 1.45 \\
\hline 
\end{tabular}
\caption{Low--lying states for  $^{129}$Sn. Energies in MeV.}
\label{tab:129sn}
\end{center}
\end{table}

\begin{table}[htbp]
\begin{center}
\begin{tabular}{cccc}
\hline 
\multicolumn{4}{c}{ $^{127}$Sn} \\ 
{$J^{\pi}_i$} &{Exp} &{$J^{\pi}_i$} &{Theory} \\
\hline 
$(11/2^{-})$         & 0.00 & $3/2^{+}_{1}$  & 0.00 \\
$(3/2^{+})$          & 0.01 & $11/2^{-}_{1}$ & 0.02 \\
$(1/2^{+})$          & 0.26 & $1/2^{+}_{1}$  & 0.09 \\
$(9/2^{+},11/2^{+})$ & 0.65 & $3/2^{+}_{2}$  & 0.86 \\
$(7/2^{+})$          & 0.81 & $9/2^{-}_{1}$  & 0.86 \\
$(3/2^{+})$          & 0.95 & $5/2^{+}_{1}$  & 0.89 \\
$(7/2^{-},9/2,11/2)$ & 0.96 & $7/2^{-}_{1}$  & 0.93 \\
$(7/2^{+})$          & 1.05 & $15/2^{-}_{1}$ & 1.11 \\
$(1/2,3/2)$          & 1.09 & $5/2^{+}_{2}$  & 1.20 \\
$(7/2^{-},9/2,11/2)$ & 1.56 & $1/2^{+}_{2}$  & 1.31 \\
$(7/2^{+})$          & 1.60 & $3/2^{+}_{3}$  & 1.34 \\
$(7/2,9/2,11/2^{+})$ & 1.91 & $7/2^{+}_{1}$  & 1.35 \\
$(7/2^{+})$          & 2.02 & $13/2^{-}_{1}$ & 1.46 \\

\hline 
\end{tabular}
\caption{Low--lying states for  $^{127}$Sn. Energies in MeV.}
\label{tab:127sn}
\end{center}
\end{table}

\begin{table}[htbp]
\begin{center}
\begin{tabular}{cccc}
\hline 
\multicolumn{4}{c}{ $^{125}$Sn} \\ 
{$J^{\pi}_i$} &{Exp} &{$J^{\pi}_i$} &{Theory} \\
\hline 
$11/2^{-}$           & 0.00 & $11/2^{-}_{1}$ & 0.00 \\
$3/2^{+}$            & 0.03 & $1/2^{+}_{1}$  & 0.04 \\
$1/2^{+}$            & 0.22 & $3/2^{+}_{1}$  & 0.04 \\
$(9/2^{-})$          & 0.62 & $9/2^{-}_{1}$  & 0.71 \\
$7/2^{+}$            & 0.86 & $3/2^{+}_{2}$  & 0.83 \\
$1/2,3/2$            & 0.93 & $5/2^{+}_{1}$  & 0.89 \\
$(7/2^{-})$          & 0.94 & $7/2^{-}_{1}$  & 0.92 \\
$7/2^{(+)}$          & 1.06 & $5/2^{+}_{2}$  & 1.06 \\
$1/2,3/2$            & 1.07 & $15/2^{-}_{1}$ & 1.07 \\
$(1/2^{+},3/2^{+},5/2^{+})$ & 1.19 & $7/2^{+}_{1}$ & 1.08 \\
$(5/2)^{+}$          & 1.26 & $1/2^{+}_{2}$  & 1.23 \\
$7/2^{+}$             & 1.36 & $3/2^{+}_{3}$  & 1.29 \\
$(5/2)^{+}$          & 1.54 & $13/2^{-}_{1}$ & 1.33 \\
\hline 
\end{tabular}
\caption{Low--lying states for  $^{125}$Sn. Energies in MeV.}
\label{tab:125sn}
\end{center}
\end{table}

\begin{table}[htbp]
\begin{center}
\begin{tabular}{cccc}
\hline 
\multicolumn{4}{c}{ $^{123}$Sn} \\ 
{$J^{\pi}_i$} &{Exp} &{$J^{\pi}_i$} &{Theory} \\
\hline 
$11/2^{-}$           & 0.00 & $11/2^{-}_{1}$ & 0.00 \\
$3/2^{+}$            & 0.03 & $1/2^{+}_{1}$  & 0.03 \\
$1/2^{+}$            & 0.15 & $3/2^{+}_{1}$  & 0.08 \\
$(9/2)^{-}$          & 0.62 & $9/2^{-}_{1}$  & 0.67 \\
$(3/2^{+},5/2^{+})$  & 0.87 & $7/2^{+}_{1}$  & 0.83 \\
$5/2^{+}$            & 0.90 & $3/2^{+}_{2}$  & 0.88 \\
$(3/2)^{+}$          & 0.92 & $7/2^{-}_{1}$  & 0.91 \\
$7/2^{-}$            & 0.93 & $5/2^{+}_{1}$  & 0.92 \\
$(7/2)^{+}$          & 1.04 & $5/2^{+}_{2}$  & 0.94 \\
$(1/2,3/2)^{+}$      & 1.07 & $15/2^{-}_{1}$ & 1.05 \\
$15/2^{-}$           & 1.11 & $1/2^{+}_{2}$  & 1.21 \\
\hline 
\end{tabular}
\caption{Low--lying states for $^{123}$Sn. Energies in MeV.}
\label{tab:123sn}
\end{center}
\end{table}

\begin{table}[htbp]
\begin{center}
\begin{tabular}{cccc}
\hline 
\multicolumn{4}{c}{ $^{121}$Sn} \\ 
{$J^{\pi}_i$} &{Exp} &{$J^{\pi}_i$} &{Theory} \\
\hline 
$3/2^{+}$            & 0.00 & $1/2^{+}_{1}$ & 0.00 \\
$11/2^{-}$           & 0.01 & $11/2^{-}_{1}$  &  \\
$1/2^{+}$            & 0.06 & $3/2^{+}_{1}$  & 0.07 \\
$(7/2,9/2)^{-}$      & 0.66 & $7/2^{+}_{1}$  & 0.65 \\
$3/2^{+},5/2^{+}$    & 0.87 & $5/2^{+}_{1}$  & 0.73 \\
$3/2^{+},5/2^{+}$    & 0.91 & $3/2^{+}_{2}$  & 0.91 \\
$7/2^{+}$            & 0.93 & $7/2^{-}_{1}$  &  \\
$(7/2)^{-}$          & 0.95 & $5/2^{+}_{2}$  & 0.94 \\
$3/2^{+},5/2^{+}$    & 1.10 & $1/2^{+}_{2}$  & 1.20 \\
$5/2^{+}$            & 1.12 & $15/2^{-}_{1}$ &  \\
$(5/2)^{+}$          & 1.15 & $3/2^{+}_{2}$  & 1.20 \\
$(7/2^{+},9/2^{+})$  & 1.35 & $7/2^{+}_{2}$  & 1.24 \\
$(5/2)^{+}$          & 1.40 & $5/2^{+}_{3}$  & 1.26 \\
\hline 
\end{tabular}
\caption{Low--lying states for  $^{121}$Sn. Energies in MeV.}
\label{tab:121sn}
\end{center}
\end{table}

Characteristic for all the odd isotopes studied in this work is that the 
$1/2^{+}_{1}$, $3/2^{+}_{1}$, $11/2^{-}_{1}$ states are nearly degenerate,
and there is then a gap to the next states located about 0.6 MeV higher. This
feature can be traced back to the single--hole spectrum where the single--hole
orbitals $1d_{3/2}$, $0h_{11/2}$ and $2s_{1/2}$ are closely degenerate and
well separated from the $1d_{5/2}$ and $0g_{7/2}$ orbitals. The three 
lowest--lying states in $^{129-121}$Sn are satisfactorily reproduced in our 
model. 
Due to the near degeneracy of these states we do not always get the states in 
the correct order, but the deviation between theory and experiment never 
exceeds 0.2 MeV.

Considering now the negative parity states, in addition to the $11/2^{-}_{1}$ 
state, which is ground state in $^{127-123}$Sn and nearly degenerate with 
the groundstate in $^{129}$Sn, there are low--lying $7/2^{-}$ and $9/2^{-}$ 
states. These states can typically be constructed by coupling an odd
number of $0h_{11/2}$ holes to an even number of $1d_{3/2}$ or $2s_{1/2}$ 
holes. To excite holes into the three orbitals mentioned here costs little 
energy, and from this simple picture it is possible to understand the low 
energies of these states.
The calculated $7/2^{-}_{1}$ state is in perfect agreement with experiment 
throughout the whole sequence of isotopes. We fail in reproducing the 
$9/2^{-}_{1}$ state in $^{129}$Sn. In $^{127}$Sn a comparison is difficult 
since the experimental $9/2^{-}_{1}$ state is not clearly identified. For the 
other isotopes the agreement between the calculated and the experimental 
$9/2^{-}_{1}$ state is good.

\section{Summary and conclusions}

A major aim of this work has been to provide a severe test of the foundation
of the effective  interaction. This is done by performing extensive shell
model calculations over a large mass region.
Most previous shell model calculations have been made for nuclei with only 
two or a few nucleons outside a closed--shell core. It is of interest to 
extend  this test to systems with several valence nucleons. For this the Sn 
isotopes are particularly suitable. From the vicinity of the doubly magic 
$^{100}$Sn to beyond the doubly magic $^{132}$Sn experimental data is now 
available. 

In this work we have investigated the range of heavy Sn isotopes between 
A=120 and A=130. These are described in terms of valence neutron holes with 
respect to the $^{132}$Sn core. By and large, the essential spectroscopic 
properties are well described in our shell model scheme. In particular, it 
is gratifying that good results are obtained for nuclei far away from the 
$^{132}$Sn core with respect to which the effective interaction and the 
single--particle energies are defined.

For the whole sequence of even isotopes the calculated $0^{+} - 2^{+}$ 
spacing is approximately constant and in almost perfect agreement with 
experiment. The other yrast states tend however to be too highly excited. 
All negative parity states are slightly to high as well. The reason seems
to be that there is too strong a coupling between holes in the 
$0h_{11/2}$ intruder orbital and holes in the other orbitals. Small 
adjustments of the $J=0$ and $2$ effective matrix elements 
$\langle n_{j}^{2}|V_{\rm eff}|0h_{11/2}^{2} \rangle _{J}$, where $n_{j}$ 
are the orbitals $0g_{7/2}$, $1d_{5/2}$, $1d_{3/2}$ and $2s_{1/2}$, give 
results in better agreement with experiment.

Our shell model scheme has been greatly successful in describing 
the odd tin isotopes. There is roughly a one--to--one 
correspondence between the calculated and experimental states below 
1.0 -- 1.5 MeV, and the states are generally in correct order. 
Usually, odd nuclei are more tricky to handle than the even nuclei, since 
they are more sensitive to the underlying structure, such as the 
single--particle energies and the interaction. However, in the present 
calculation we get at least as satisfactory results for the odd as 
for the even isotopes.

In summary we may conclude that the relative location of the states is
satisfactory, but it ought to be mentioned that the absolute values of the
interaction energies are far off. As moving away from the closed Z=50, 
N=82 core, the systems become too strongly bound compared to experiment. 
Further investigations of this problem are necessary. Left to be studied 
in more detail is the influence of three--body forces \cite{eh97}. We 
would like to find out whether the three--body contributions will be of 
significant importance as the number of valence particles grows, and thus 
will bring the theoretical binding energies closer to the experimental 
values. Another explanation could be that the problems with reproduction 
of the binding energies may be ascribed to the two--body effective 
interaction itself and thus this ought to be subjected to further tests 
before definite conclusions can be drawn.

This work was initiated while one of us, M.H.J., was at the
European Centre for Theoretical studies in Nuclear Physics and Related Areas,
Trento, Italy. 
Financial support from the Instituto Trentino di Cultura, 
Trento, and the Research
Council of Norway (NFR) is greatly acknowledged. The calculations have been 
carried out at the IBM cluster at the University of Oslo. Support
for this from the NFR is acknowledged.


\begin{thebibliography}{200}
\bibitem{bf84} J.\ Blomqvist and B.\ Fogelberg, Phys.\ Lett.\ 
B 137 (1984) 20.
\bibitem{ehho95}T.\ Engeland, M.\ Hjorth-Jensen, A.\ Holt and
E.\ Osnes,
Phys.\ Scripta T56 (1995) 58.
\bibitem{ehho93}T.\ Engeland, M.\ Hjorth-Jensen, A.\ Holt and
E.\ Osnes,
Phys.\ Rev.\ C 48 (1993) R535.
\bibitem{heho94a}A.\ Holt, T.\ Engeland, M.\ Hjorth-Jensen and
E.\ Osnes,
Nucl.\ Phys.\  A 570 (1994)  137c.
\bibitem{hko95}  M.\ Hjorth-Jensen, T.\ T.\ S.\ Kuo and
E.\ Osnes, Phys.\ Reports 261 (1995) 125.
\bibitem{eh97} T.\ Engeland and M.\ Hjorth--Jensen, in preparation.
\bibitem{nicu94} N.\ Sandulescu, J.\ Blomqvist and R.J.\ Liotta,
Nucl.\ Phys.\ A 582 (1994) 257; Physica Scripta T56 (1995) 84.
\bibitem{brown94} B.A.\ Brown and K.\ Rykaczewski, 
Phys.\ Rev.\ C 50 (1994) R2270.
\bibitem{nicu93} A.\ Insolia, N.\ Sandulescu, J.\ Blomqvist and
R.J.\ Liotta, Nucl.\ Phys.\ A 550 (1992)  34. 
\bibitem{pomar90} C.\ Pomar, J.\ Blomqvist, R.J.\ Liotta and
A.\ Insolia, Nucl.\ Phys.\ A 515 (1990) 381.
\bibitem{cdbonn} R.\ Machleidt, F.\ Sammarruca and Y.\ Song,
Phys.\ Rev.\ C 53 (1996)
\bibitem{nim} V.G.J.\ Stoks, R.A.M.\ Klomp, C.P.F. Terheggen and J.J.\
de Swart, Phys.\ Rev.\ C 48 (1993) 792.
\bibitem{v18} R.B.\ Wiringa, V.G.J.\ Stoks and R.\ Schiavilla, Phys.\ Rev.\
C 51 (1995) 38.
\bibitem{mac89}  R.\ Machleidt, Adv.\ Nucl.\ Phys.\ 19 (1989)  189. 
\bibitem{ko90}  T.T.S.\ Kuo and E.\ Osnes, Folded-Diagram Theory of the
Effective Interaction in Atomic Nuclei, Springer Lecture Notes in Physics,
(Springer, Berlin, 1990) Vol.\ 364.
\bibitem{whit77} R.R.\ Whitehead, A.\ Watt, B.J.\ Cole and I.\
Morrison, Adv.\ Nucl.\ Phys.\ 9 (1977) 123.
\bibitem{nndc} National Nuclear Data Center, Brookhaven National Laboratory,
Upton, N.Y., USA, http://www.nndc.bnl.gov/.
\bibitem{holt97} A.\ Holt, T.\ Engeland, M.\ Hjorth--Jensen, E.\ Osnes and
J.\ Suhonen, Nucl.\ Phys.\ A618 (1997) 107.
\bibitem{fiel77} H.\ Fielding {\em et al.}, Nucl.\ Phys.\ A281 (1977) 389.
\bibitem{bron79} J.\ Bron {\em et al.}, Nucl.\ Phys.\ A318 (1979) 335.
\bibitem{talm77} I.\ Talmi, Proceedings of the International School of 
Physics "Enrico Fermi", Elementary Modes of Excitation in Nuclei, 
(North--Holland Publishing Company, Amsterdam, 1977) 352. 
\bibitem{kiss60} L.\ S.\ Kisslinger and R.\ A.\ Sorensen, Mat.\ Fys.\ Medd.\
Dan.\ Vid.\ Selsk.\ 32, no. 9 (1960).
\bibitem{foge84} B.\ Fogelberg and J.\ Blomqvist, Phys.\ Lett.\ 137B (1984) 
20.
\bibitem{brod92} R.\ Broda {\em et al.} Phys.\ Rev.\ Lett.\ 68 (1992) 1671.
\bibitem{hjor97} M.\ Hjorth--Jensen, unpublished.
\end{thebibliography}
\end{document}